\documentclass[prl,aps,twocolumn,floats,showpacs]{revtex4}

\newif\ifpdf\ifx\pdfoutput\undefined\pdffalse\else\pdfoutput=1\pdftrue\fi

\ifpdf  \usepackage[pdftex]{graphics}
                                 \usepackage{epstopdf}
  \else\usepackage{graphics}\fi

\usepackage{graphicx}
\usepackage{epsfig}
\usepackage{psfrag}

\usepackage{amsfonts}
\usepackage{amsmath}
\usepackage{amsthm, amssymb}

\newlength{\figwidth} 
\newcommand{\fig}[3] 
{
\begin{figure}[!tb] 
\vspace*{-0.1cm} 
\[ 
\includegraphics[width=\figwidth]{#1} 
\] 
\vspace{-0.3cm}
\caption{\label{#2} 
\small#3 
\vspace{-0.3cm}
}
\end{figure}}

\newcommand{\ve}[1]{\mathbf{#1}}
\newcommand{\kk}{\ve{k}}

\newcommand{\QQ}{\ve{Q}}

\newcommand{\vS}{\mathbf{S}}

\newcommand{\vM}{\vec{M}}

\newcommand{\up}{\uparrow}
\newcommand{\dn}{\downarrow}

\newcommand{\beq}{\begin{equation}}
\newcommand{\eeq}{\end{equation}}
\newcommand{\non}{\nonumber}
\newcommand{\hx}{\hat{x}}
\newcommand{\hy}{\hat{y}}

\newcommand{\BaCo}{Ba(Fe$_{1-x}$Co$_x$)$_2$As$_2\,$}
\newcommand{\CaCo}{Ca(Fe$_{1-x}$Co$_x$)$_2$As$_2\,$}
\newcommand{\BaFe}{BaFe$_2$As$_2\,$}
\newcommand{\CaFe}{CaFe$_2$As$_2\,$}

\begin{document}
\title{Interplay of orbital and spin ordering in the iron pnictides}
\author{Andriy H. Nevidomskyy}
\affiliation{Department of Physics and Astronomy, Rice University, TX 77005, USA}

\date{\today}
\begin{abstract}
A number of recent experiments exhibit electronic anisotropy in the iron pnictides, and there is a growing body of experimental evidence that its origin is related to orbital ordering in Fe $d_{xz}$ and $d_{yz}$ orbitals. We examine this problem in the parent compounds of the iron pnictides by a combination of \emph{ab initio} band theory calculations, phenomenological Ginzburg--Landau theory of coupled orbital and magnetic order parameters, and a microscopic mean-field study of the Kugel--Khomskii model. 
We find that orbital ordering  is sufficient to explain a number of key experimental observations, in particular linear correlation between the orthorhombic lattice distortion and the magnetic ordered moment.
We predict that orbital polarization should scale as a square of magnetic moment close to $T_N$. Mediated by orbital polarization, the effective spin-spin exchange interactions develop anisotropy in the $ab$-plane, in accord with recent neutron scattering measurements. 
\end{abstract}


\maketitle


Proximity of the superconducting phase to antiferromagnetic (AFM) order in the phase diagram of the newly discovered iron-based superconductors, similar to the cuprates family, has fueled renewed interest in the role of magnetism in unconventional superconductivity. Understanding the nature of magnetic order in the iron pnictides is an important aspect of these materials, and has received much attention 
\cite{delaCruz08, Huang08,Kaneko08, Kumar08, Zhao09, Diallo09}. 
Intriguingly, the magnetic order 
is preceded or coincides with the tetragonal to orthorhombic structural transition ~\cite{delaCruz08, Huang08, Kumar08}, breaking the $C_4$ lattice symmetry.
A phase with spontaneously broken rotational symmetry, often referred to as \emph{nematic} phase, has also been recently reported in the pseudogap phase of high-temperature cuprate superconductors~\cite{Daou10, Lawler10} raising the question of its role in superconductivity (SC). In the iron pnictides, the nematicity was observed in neutron scattering~\cite{Zhao09} and scanning tunneling microscopy~\cite{Chuang10} 
below the magnetic ordering temperature $T_N$.
It is also seen in the resistivity anisotropy measured in detwinned \BaCo crystals~\cite{Fisher-detwin, Tanatar-detwin} even 
above $T_N$.
A recent observation of highly anisotropic spin-wave dispersion above $T_N$ in the parent compound \BaFe was attributed to spontaneous ``spin nematicity''~\cite{Harriger10}. Similar anisotropy was also found inside the superconducting phase of optimally doped \BaCo~\cite{Hayden10}.

These observations raise the question of the role of broken $C_4$ symmetry and its relation to the magneto-structural phase transition in the iron pnictides. Two distinct scenarios have been proposed theoretically. 
The first concept, ``spin nematic'', is based on the spontaneously broken $Z_2$ Ising symmetry between two collinear ordering wave-vectors $\QQ_a=(\pi,0)$ and $\QQ_b=(0,\pi)$ chosen by ``order from disorder'' mechanism~\cite{Fang08, Xu08, Yildrim08}. 
It was first proposed by Chandra, Coleman and Larkin that such Ising spin symmetry can be spontaneously broken at a temperature higher than the magnetic ordering $T_N$~\cite{CCL}, where the ordered moment $\langle M_i\rangle=0$, but the Ising variable $\sigma\propto \langle \vM_A\cdot \vM_B\rangle = \pm1$ acquires an expectation value ($A$ and $B$ are two sublattices). 
This mechanism, although breaks the lattice $C_4$ symmetry,  does not however imply a broken $SU(2)$ spin-rotational symmetry. In this respect, it differs from the term ``nematic'' originally used in studies of liquid crystals and could be perhaps more aptly referred to as ``spin-Ising ordering'', following Ref.~\onlinecite{CCL}.

The second scenario proposes that the broken orthorhombic symmetry stems from unequal population of the $d_{xz}$ and $d_{yz}$ orbitals, resulting in the so-called ferro-orbital ordering~\cite{Singh09, Lee09, Lv09, Chen09, Chen10, Lv10}. The order parameter here is the orbital polarization $P=\langle n_{xz}-n_{yz}\rangle$, which explicitly breaks the $C_4$ symmetry. The state with $P\neq0$ could be called an ``electron nematic'' and closely corresponds to the nematic order proposed for the bilayer ruthenate Sr$_3$Ru$_2$O$_7$~\cite{Sr327} under applied magnetic field. The orbital ordering has gained support from the recent quadrupole resonance~\cite{NQR-1111} and ARPES measurements~\cite{Yi-ARPES}.

Despite very different physical origin, both mechanisms allow for linear coupling between the structural orthorhombic distortion, $\delta = \frac{a-b}{a+b}$ and the respective order parameter, and cannot be distinguished on symmetry grounds alone. 
%
In this work, we report the combined study based on 
density functional theory, Ginzburg--Landau expansion and microscopic theory, all lending strong support to the orbital ordering in the pnictides.

The \emph{ab initio} density functional theory (DFT) calculations have been performed on the parent compound \BaFe using the full-potential augmented plane-wave basis, as implemented in the WIEN2k code~\cite{Wien2k}, with the generalized gradient approximation (GGA) for the exchange-correlation functional~\cite{PBE}. Since we are interested in the spontaneously broken $C_4$ symmetry, we have adopted the experimental tetragonal structure~\cite{Ba122-structure}, to avoid the effects of the orthorhombic lattice distortion which explicitly breaks the $90^{\circ}$ rotation symmetry. For the $\kk$-point sampling, the $6\times 6\times 3$ mesh was used in the reduced part of the Brillouin zone corresponding to the 4-atom antiferromagnetic unit cell (4 f.u./cell). The magnetic structure is collinear with ordering wave-vector $(\pi,0)$ or $(0,\pi)$ in the 1-Fe unit cell notation. The calculated ordered moment $g M_{tot}$=1.91 $\mu_B$ is more than twice larger than the measured value $gM_{exp}=0.91 \mu_B$, a known problem of the DFT in the iron pnictides~\cite{Mazin08}. One can attempt to include the effect of Coulomb repulsion in the LDA+$U$ framework~\cite{Anisimov91}, however usually this exacerbates the problem. Recently, it was shown however~\cite{Cricchio-lowspin, Machida-lowspin} that a low-spin configuration of Fe can be stabilized within LDA+$U$ approach, in agreement with the tight-binding calculations~\cite{Bascones-lowspin}. We indeed find a low-moment solution with $gM_{low}=0.57\mu_B$/Fe within the so-called ``around mean field'' GGA+$U$ scheme~\cite{AMF}, using the values of $U=2.7$~eV and Hund's $J=0.79$~eV calculated from a constrained-RPA scheme~\cite{Aichhorn09} for LaFeAsO~\cite{remark-Aichhorn}. The low-moment state is stabilized by a significant energy 0.22~eV per Fe atom compared to the high-moment state.

\fig{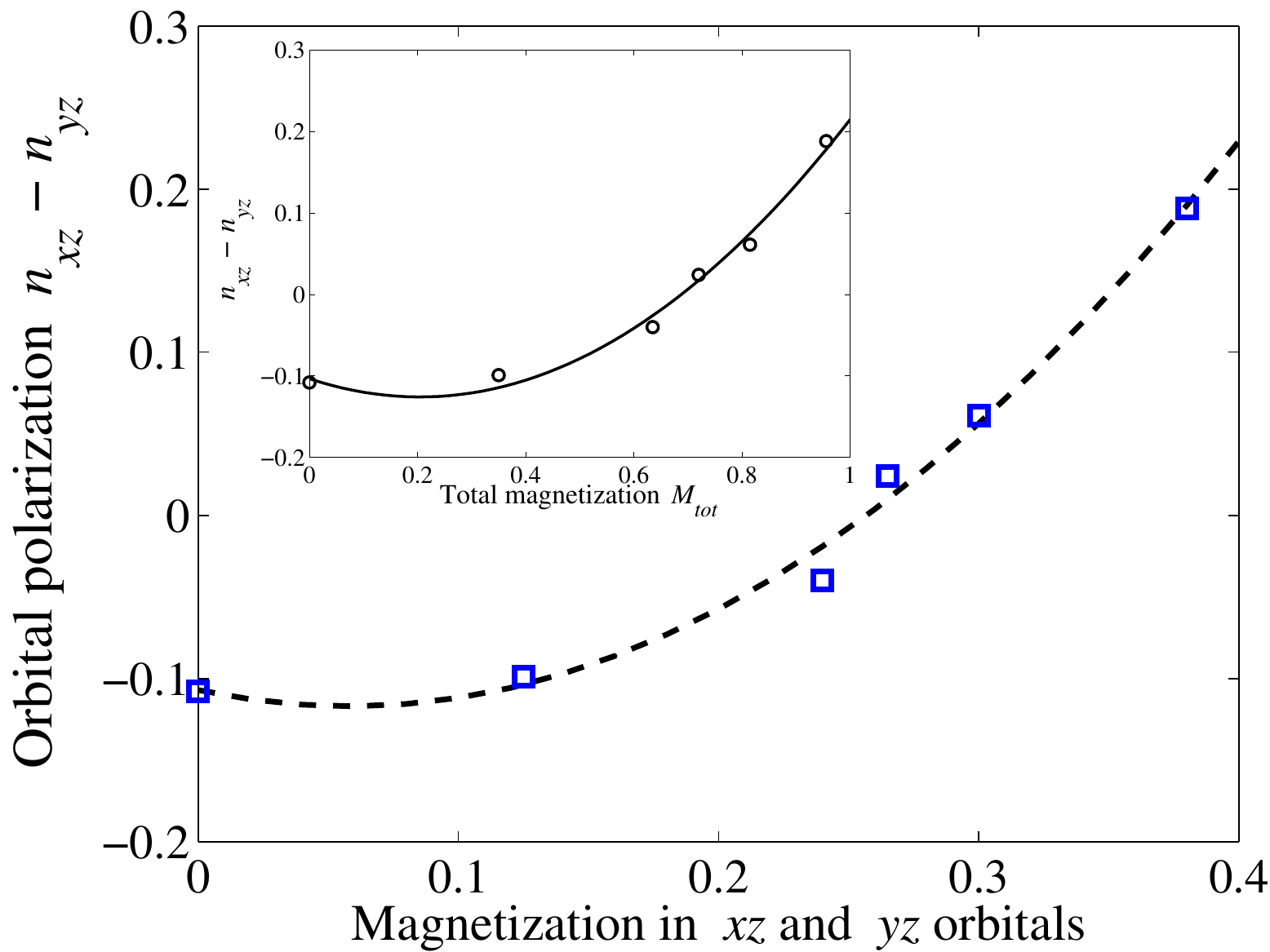}{Fig.DFT}{Orbital polarization $P=n_{xz}-n_{yz}$ as function of (partial) ordered moment $M=\frac{1}{2}(n_{xz\up} - n_{xz\dn} + n_{yz\up} -n_{yz\dn})$.  The line is a quadratic fit to the data: $P=P_0 + c_P M^2$. The inset shows $P$ plotted vs. total magnetization $M_{tot}$ on the Fe site and follows a similar quadratic dependence.\vspace{-3mm}}

In both the GGA and GGA+$U$ calculations, we find a non-zero value of orbital ordering: $P_\text{GGA}=0.19$ and $P_\text{GGA+U}=0.30$ in the low-moment $(\pi,0)$ phase. The GGA calculated polarization agrees well with the value reported by the recent ARPES measurement~\cite{Yi-ARPES} on \BaCo.
In order to elucidate the dependence of orbital polarization $P$ on the size of the ordered moment $M$, we have performed a series of calculations effectively suppressing the spin moment by introducing an orbital potential that couples to $L^z$~\cite{Eriksson-orbital}. Shown in  Fig.~\ref{Fig.DFT} is the resulting orbital polarization $P=n_{xz}-n_{yz}$ plotted vs. ordered moment in Fe $d_{xz}$ and $d_{yz}$ orbitals $M=\frac{1}{2}(n_{xz\up} - n_{xz\dn} + n_{yz\up} -n_{yz\dn})$.  The data points can be fitted by a quadratic dependence $P=P_0 + c_P M^2$, which also holds as a function of the total ordered moment $M_{tot}=\frac{1}{2}(n_\up - n_\dn)$ on Fe site (inset of Fig.~\ref{Fig.DFT}).

Phenomenologically, 
the inteplay between orbital and spin polarization
can be captured by an effective Ginzburg--Landau theory with two order parameters:
\begin{eqnarray}
F[M,P] &=& \left( \alpha P^2 + \frac{u}{4}P^4 - w P\delta\right) - \gamma P^2 M^2 \nonumber \\
&+& \left(rM^2 + \frac{v}{4}M^4\right) + \dots  \label{F}
\end{eqnarray}
where $r=\frac{T-T_N}{T_N}$ is the reduced temperature of the AFM transition and $\delta$ is the orthorhombic lattice distortion. 
We consider the situation below the orbital ordering transition temperature, $T<T_O$, so that the reduced temperature $\alpha = \frac{T-T_O}{T_O}<0$.
The choice of the biquadratic coupling $-\gamma P^2M^2$ is warranted by the microscopic Kugel-Khomskii theory (see below). 
Let us first consider the case $\delta=0$. The saddle-point solution of Eq.~(\ref{F}) yields:
\begin{eqnarray}
M\! &=&\! \sqrt{\frac{2(-r + \gamma P^2)}{v}}, \quad \text{if }  r-\gamma P^2 <0 \label{M}\\
P\! &=&\! \sqrt{\frac{-2\alpha}{u}\left(1 + \frac{\gamma}{|\alpha|}M^2 \right)} \approx P_1 + c_P M^2 + \mathcal{O}(M^4) \label{P},
\end{eqnarray}
with the coefficient $c_P(T)\! =\! \frac{\gamma}{2|\alpha|} P_0 \left(1+\frac{\gamma}{|\alpha|}M_0^2\right)^{-1/2} $ and $P_1(T)\!=\!P_0\sqrt{1+\frac{\gamma}{|\alpha|}M_0^2}$ expressed through the mean-field values $M_0(T)=\sqrt{-2r/v}$ and $P_0(T)=\sqrt{-2\alpha/u}$ in the absence of coupling between the two order parameters.

We see that the Landau theory (\ref{P}) naturally accounts for quadratic dependence of orbital polarization on the size of the ordered moment, as obtained earlier from \emph{ab initio} calculations (Fig.~\ref{Fig.DFT}).
It also follows from Eq.~(\ref{M}) that the magnetic ordering temperature becomes higher due to coupling to orbital ordering: $T_N^* = T_N(1+2|\alpha^*|\gamma/u)$ with $|\alpha^*|=(T_O - T_N^*)/T_O>0$.

Let us now consider the case of non-zero orthorhombic distortion $\delta=(a-b)/(a+b)$. Due to coupling of distortion to the orbital ordering in Eq.~(\ref{F}), we obtain $w\delta = -2\alpha P + u P^3 - 2\gamma P M^2$. This results in the leading linear contribution to orbital order parameter $P\sim \delta + \mathcal{O}(\delta^3)$ for temperatures not too far from $T_O$. It follows from Eq.~(\ref{M}) that the magnetization 
\beq
M^2 = \frac{-2r}{v} + \frac{2\gamma w^2}{v}\delta^2 + \mathcal{O}(\delta^4).
\eeq
In particular near $T_N$, where the first term is negligible, magnetization scales linearly with orthorhombic distortion, $M(\delta)\propto \delta$. This prediction of the Landau theory is entirely consistent with the neutron scattering measurements on La(O$_{1-x}$F$_x$)FeAs~\cite{delaCruz08}, SrFe$_2$As$_2$~\cite{Kaneko08, Li09} and \CaCo~\cite{Prokes11}. 

Note that instead of biquadratic term $-\gamma P^2M^2$, a  coupling of the form $-\kappa P M^2$ is also allowed by symmetry, however the analysis shows that it would result in the quadratic dependence of structural distortion on magnetic ordered moment, $w \delta = -2\alpha P + uP^3 -\kappa M^2 = C_1 + C_2M^2 + \mathcal{O}\left(M^4\right)$.
This would clearly contradict the aforementioned neutron scattering measurements.

Intriguingly, the \emph{ab initio} calculations find that orbital polarization is split unequally between the two spin components. Both in GGA and GGA+$U$, the largest contribution comes from the minority spin component. In other words, the expectation value of the spin-antisymmetric orbital polarization $P_s=P_\up - P_\dn$ is non-zero, as well as total $P=P_\up+P_\dn$. While $P$ only breaks $C_4$ symmetry, $P_s$ also break the time-reversal symmetry $T$ (but preserves the product $T\times C_4$). 
It was proposed in the context of the nematic phase in Sr$_3$Ru$_2$O$_7$ that appearance of both $P$ and $P_s$ could lead to an additional symmetry-allowed term in the Ginzburg--Landau free energy $\propto M(P P_s)=M(P_\up^2 - P_\dn^2)$~\cite{Raghu-nematic09}. Our analysis shows however that if it were the case, in the presence of orthorhombic lattice distortion $P_\sigma \propto \delta$ and hence $M\propto P_\up^2 - P_\dn^2 \propto \delta^2$, instead of the experimentally observed $M\propto \delta$.

In order to provide a more microscopic basis for the Ginzburg--Landau theory Eq.~(\ref{F}), we studied the interplay between orbital and spin physics within the framework of the Kugel--Khomskii model~\cite{KK}. 
In the $t_{2g}$ orbital basis, the Kugel--Khomskii model 
can be schematically written as follows~\cite{Ishihara04, Kruger09}:
\begin{eqnarray}
H\!\! &=& \!\!J_1\sum_i\left[(\vS_i\cdot \vS_{i+\hx}+1)\tau_i^a \tau_{i+\hx}^a + (\vS_i\cdot \vS_{i+\hy}+1)\tau_i^b \tau_{i+\hy}^b \right] \nonumber \\
& +& J_2\sum_{\langle\langle i,j\rangle\rangle} (\vS_i\cdot \vS_j) - J_d\sum_{\langle i,j\rangle} (\vS_i\cdot \vS_j),
\label{KK}
\end{eqnarray}
where $\vS_i=1$ is the magnetic moment localized on Fe $d_{xz}$ and $d_{yz}$ orbitals, and $\{\tau^a,\tau^b\}$ are the pseudospin operators that act in the orbital subspace of $|xz\rangle$ and $|yz\rangle$ states and depend on the directionality of the Fe-Fe bond. The last term describes the direct (ferromagnetic) exchange between Fe spins, proposed by R. Singh~\cite{Singh09}.

The key question is the occupation of the $d_{xz}$ and $d_{yz}$ orbitals. 
In Ref.~\cite{Kruger09}, a similar model was derived for the configuration with total of 2 electrons in three $t_{2g}$ orbitals, whose spins are aligned to form spin $S=1$. However our  GGA and GGA+$U$ calculation predict that the total number of electrons in the $t_{2g}$ orbitals is about 3.16 in the paramagnetic phase, with each orbital equally populated close to half filling.  This agrees with the Fe$^{2+}$ multiplet structure calculated from LDA+DMFT~\cite{Haule-JHunds}. Thus the Fe$^{2+}$ ground state appears to lie in the $t_{2g}^3$ rather than $t_{2g}^2$ sector, as was assumed in Ref.~\onlinecite{Kruger09}. Nevertheless, if one only considers $|xz\rangle$ and $|yz\rangle$ states, they can still be described by total spin $S=1$ and Eq.~(\ref{KK}) should hold. However the crucial difference from earlier treatments~\cite{Singh09, Kruger09, Chen09} is that $\tau_i$ are \emph{not} Ising variables, since both $xz$ and $yz$ orbitals are partially occupied.

In what follows, we develop the mean-field treatment of the model (\ref{KK}), approximating the orbital average $\langle \tau_i^a \tau_{i+\hx}^a \rangle  \sim \left\langle  (\tau_i^a)^{2} \right\rangle \approx n_{xz}^2$ and similarly $\langle \tau_i^a \tau_{i+\hx}^a \rangle \approx n_{yz}^2$. Eq.~(\ref{KK}) can be decomposed into a sum of orbital and spin parts: $H_\text{MF} = E_0 + H_\text{o}+H_\text{sp} $, where
\begin{eqnarray}
-E_0\!\! &=&\!\! J_1\sum_i\!\left[\langle\vS_i\cdot \vS_{i+\hx}+1\rangle n_{xz}^2+ \langle\vS_i\cdot \vS_{i+\hy}+1\rangle n_{yz}^2 \right]\label{E0}\\
H_\text{o}\!\! &=&\!\! J_1\!\sum_i\langle\vS_i\cdot \vS_{i+\hx} + 1\rangle\tau_i^a \tau_{i+\hx}^a 
+ \langle\vS_i \cdot \vS_{i+\hy} + 1\rangle \tau_i^b \tau_{i+\hy}^b \non \\
H_\text{sp}\!\! &=&\!\! \sum_i\left[J_{1a}\, \vS_i \cdot \vS_{i+\hx} + J_{1b}\,\vS_i\cdot \vS_{i+\hy}\right] + J_2\!\!\sum_{\langle\langle i,j\rangle\rangle}\!\! \vS_i\! \cdot\! \vS_j \label{Hsp}
\end{eqnarray}
We see that the mean-field ground state energy takes on the form $E_0\sim -\gamma \langle \vS_i \vS_j \rangle n_\alpha^2$, justifying the choice of the biquadratic coupling in the Landau theory Eq.~(\ref{F}).

It follows from $H_\text{o}$ above that the strength of the intersite orbital coupling is proportional to $\langle \vS_i\cdot \vS_j + 1 \rangle$, i.e. the deviation of spins from the classical N\'eel state. 
This is a general feature of the Kugel--Khomskii model, well-known for the $e_g$ case~\cite{KK-Feiner, KK-Khaliullin}. The classical N\'eel state with ordering wave-vector $(\pi,\pi)$ would have infinite orbital degeneracy corresponding to $SU(2)$ pseudospin invariance. However for the case of the collinear $(\pi,0)$ order observed in the pnictides, only the orbital correlations along the antiferromagnetic $a$-axis will be massless. These massless local orbital fluctuations will affect the spin sector, reducing the staggered moment~\cite{KK-Feiner, KK-Khaliullin}.

In the effective spin Hamiltonian (\ref{Hsp}), the exchange ``constants'' depend on the orbital polarization $P$:
\begin{eqnarray}
J_{1a} &=& J_1\langle\tau_i^a \tau_{i+\hx}^a \rangle - J_d \;\approx\; J_1 (n+P)^2/4 - J_d\\
J_{1b} &=& J_1\langle\tau_i^b \tau_{i+\hy}^b \rangle - J_d \;\approx\; J_1 (n-P)^2/4 - J_d,
\end{eqnarray}
where we denoted the total number of electrons in $d_{xz}$ and $d_{yz}$ orbitals by $n=n_{xz}+n_{yz}$. From our GGA and GGA+$U$ calculations, we find $n\approx 2.5$ in the collinear ordered state, in difference to Ref.~\onlinecite{Chen09} where it was assumed that orbital degrees of freedom were Ising and that $n=1$.

At the mean-field level, we can express \mbox{$\langle \vS_i\cdot\vS_j\rangle$} through the ordered moment $M$, e.g.: $\langle \vS_i\cdot\vS_{i+\hy}\rangle = \pm M^2$, where the ``$+$'' sign corresponds to ordering wave-vector $\ve{Q}=(\pi,0)$ and ``$-$'' sign to $\ve{Q}=(0,\pi)$.
The orbital polarization is found by minimizing the ground state energy $E_0$ in Eq.~(\ref{E0}) with respect to $n_{xz}=(n+P)/2$:
\beq
n_{xz} = \frac{n}{2}(1\pm M^2), \quad P = \left\{ \begin{array}{cc}
 nM^2, &\text{for } \ve{Q}=(\pi,0) \\
-nM^2, &\text{for } \ve{Q}=(0,\pi)
\end{array} \right. \label{Pmf}
\eeq
The above two solutions are degenerate and the system chooses one of them by breaking the $Z_2$ Ising symmetry, as in the ``spin nematic'' scenario~\cite{CCL}.

Within our \emph{ab initio} GGA calculations, the ordered moment in the $xz$ and $yz$ orbitals  
$M=0.38$.
Using $n_\text{GGA}=2.54$, Eq.~(\ref{Pmf}) yields the mean-field value for orbital polarization $P_{MF}=0.37$, which is significantly higher than $P_\text{GGA}=0.19$ from the \emph{ab initio} calculation. This is not unexpected since orbital fluctuations will tend to lower the polarization from its mean field value.

We now return to the effective spin Hamiltonian, Eq.~(\ref{Hsp}). At the mean field level, the anisotropy of the effective exchange ``constants'' in $ab$-plane becomes
\beq
\eta\equiv\frac{J_{1a} - J_{1b}}{J_{1a} + J_{1b}} = \frac{2 n P}{n^2 + P^2 - 4|J_d|/J_1} \label{eta}
\eeq
Using the ARPES estimate~\cite{Yi-ARPES} $P\approx 0.18$ (consistent with our $P_\text{GGA}=0.19$), $n\approx 2.5$, and taking a realistic estimate on $|J_d|=10$~meV, $J_1=50$~meV~\cite{Harriger10}, we arrive at the value $\eta\approx 0.16$. This would suggest that $J_{1b}\approx 1.35 J_1 - J_d$, which is clearly antiferromagnetic, unless one assumes an unrealistically large value for $J_d\gtrsim J_1$.
It has been proposed~\cite{Singh09, Chen09} that orbital order coupled to Fe spins as in Eq.~(\ref{Hsp}) could explain the significant anisotropy in the spin-wave dispersion  observed by neutron scattering in \CaFe and \BaFe~\cite{Zhao09, Harriger10}. Fitting neutron spectra to the linear spin-wave theory actually results in a ferromagnetic $J_{1b}\approx-6$~meV~\cite{Zhao09}. However in Eq.~(\ref{eta}), even for a large value of $|J_d|=J_1/2$, the upper bound on $\eta$ is $\sim 0.21$, and $J_{1b}$ remains positive. 

Using the mean-field expression for orbital polarization Eq.~(\ref{Pmf}), we can express the parameters of the spin Hamiltonian in terms of the ordered moment as follows:
\beq
J_{1a/b} = \left(\frac{J_1 n^2}{4} - J_d \right) \pm \frac{J_1n^2}{2}M^2 + \mathcal{O}(M^4) \label{JviaM}
\eeq
What is striking is that the values of the exchange ``constants'' depend on the ordered moment. This has two consequences: first, the anisotropy in the spin-wave dispersion grows in the ordered phase, in turn reinforcing the orbital polarization $P\propto M^2$. Secondly, the term $\sim M^2$ in Eq.~(\ref{JviaM}) introduces quartic coupling between neigbouring spins of the form $(\vS_i\cdot\vS_j)(\vS_i\cdot\vS_j)$, making the model qualitatively different from a simple anisotropic Heisenberg model. This strongly suggests that attempts to fit the neutron spin-wave spectra to a simple $J_{1a}-J_{1b}-J_2$ model with \emph{constant} exchange parameters~\cite{Zhao09, Harriger10} are problematic, since they neglect to take quartic coupling between the spins into account.

Present work suggests that orbital ordering, as seen in our \emph{ab initio} calculations and in experiment~\cite{NQR-1111, Yi-ARPES}, is consistent with predictions of microscopic theory and explains a number of key experimental features, in particular the linear correlation of ordered moment with lattice distortion, $M\sim \delta$. We predict that orbital polarization should scale as a square of magnetic moment close to $T_N$, a feature that can be tested experimentally. We propose that the spin-wave dispersion in the magnetically ordered phase should be fitted with the moment-dependent, rather than constant, exchange parameters $J_{1a}, J_{1b}$ given by Eq.~(\ref{JviaM}).
Although self-sufficient, orbital polarization can in principle co-exist with other proposed mechanisms for spontaneously broken $C_4$ symmetry, such as spin-Ising ordering~\cite{Xu08,Fang08} or Pomeranchuk instability~\cite{Zhai09}. 

The author would like to thank E. Abrahams, R. Fernandes, I. Eremin, A. Chubukov, P. Goswami, S. Raghu, Q. Si, E. Dagotto, R. McQueeny, I. Fisher and R. Singh for valuable discussions, and to M. Machida for bringing Ref.~\onlinecite{Cricchio-lowspin} to our attention.



\vspace{-5mm}

\bibliographystyle{apsrev}
\bibliography{pnictides}

\begin{thebibliography}{48}
\bibitem[{rem()}]{remark-Aichhorn}
\bibinfo{note}{While the values of $U$ and $J$ may be different in
  BaFe$_2$As$_2$ compared to LaFeAsO, it is not essential for the discussion,
  in view of the approximate nature of LDA+$U$}.
\end{thebibliography}


\end{document}